\documentstyle[11pt,epsf]{article}

\newcommand{\np}[3]{{\sl Nucl. Phys.} {\bf #1} (19#2)~#3}
\newcommand{\pl}[3]{{\sl Phys. Lett.} {\bf #1} (19#2) #3}
\newcommand{\pr}[3]{{\sl Phys. Rev.} {\bf #1} (19#2) #3}

\newcommand{\vj}[4]{{\sl #1~}{\bf #2} (19#3) #4}

\def\be{\begin{equation}}
\def\ee{\end{equation}}
\def\bea{\begin{eqnarray}}
\def\eea{\end{eqnarray}}

\newcommand{\bfi}[1]{\begin{figure}[#1]}
\newcommand{\efi}{\end{figure}}
\newcommand{\bpi}[2]{\begin{picture}(#1,#2)}
\newcommand{\epi}{\end{picture}}

\newcommand{\nn}{\nonumber\\}

\newcommand{\prop}{\Delta}

\newcommand{\dsl}{{\not \! \partial}}
\renewcommand{\d}{\partial}
\newcommand{\T}{{\mathrm T}}
\newcommand{\Q}{{\mathrm Q}}
\newcommand{\B}{{\mathrm B}}
\newcommand{\A}{{\mathrm A}}

\newcommand{\F}{{\mathrm F}}
\newcommand{\OO}{{\mathcal O}}

\begin{document}

\pagestyle{empty}
{\hfill \parbox{6cm}{\begin{center} 
	UG-FT-91/98  \\ 
	hep-ph/9808315 \\
	August 1998                   
\end{center}}} 

\renewcommand{\thefootnote}{\fnsymbol{footnote}}
\vspace*{1cm}
\begin{center}
\large{\bf Differential renormalization of gauge 
theories\footnote{Presented at the Zeuthen Workshop on Elementary
Particle Physics ``Loops and Legs in Gauge Theories'', Rheinsberg,
Germany, April 19-24, 1998.}}
\vskip .6truein
\centerline {\large{\sc F. del \'Aguila and M. P\'erez-Victoria}}
\end{center}
\vspace{.3cm}
\begin{center}
{Dpto. de F\'{\i}sica Te\'orica y del Cosmos,  
 Universidad de Granada, \\
 18071 Granada, Spain} 
\end{center}
\vspace{1.5cm} 
 
\centerline{\bf Abstract} 
\medskip 

The scope of constrained differential renormalization is to provide
renormalized expressions for Feynman graphs, preserving at the same
time the Ward identities of the theory. It has been shown recently
that this can be done consistently at least to one loop for abelian
and non-abelian gauge theories. We briefly review these results,
evaluate as an example the gluon selfenergy in both coordinate and
momentum space, and comment on anomalies. 

\vspace*{1.5cm}

\pagestyle{plain}
\renewcommand{\thefootnote}{\arabic{footnote}}
\setcounter{footnote}{0}

\section{Introduction}
Differential renormalization (DR) is a renormalization method in
coordinate space which provides finite Green functions without any
intermediate regulator or counterterms~\cite{FJL}. It has been
applied to a variety of problems (a complete list of references can
be found in Ref.~\cite{techniques}). As initially formulated, DR is
too general for it introduces many arbitrary constants in the process
of renormalization, which have to be fixed at the end of the
calculations requiring the fulfilment of the relevant Ward identities. 
A {\em constrained} version has been recently proposed
which aims to fix this arbitrariness from the beginning and
automatically fulfil the Ward
identities~\cite{techniques,CDR,polonia}. In Section~2 we briefly
review this constrained procedure. It has been explicitly shown to
work at one loop for gauge theories (QED, Scalar QED and QCD are
discussed in Refs.~\cite{CDR,polonia}, \cite{techniques} and~\cite{QCD},
respectively). The method is defined in coordinate space, but one can
perform the calculations also in momentum space, using the Fourier
transforms of the  renormalized expressions in coordinate space. A
computer code doing all operations automatically 
is available~\cite{program}. 
For illustration, we evaluate in Section~3
the one-loop gluon selfenergy in both coordinate and momentum space,
and verify that it is transverse, as required by gauge 
invariance~\cite{ST}.
Finally we comment on anomalies in Section~4 and 
on future prospects in the conclusions.

\section{Overview of constrained differential renormalization}
Constrained differential renormalization (CDR) is essentially a set
of rules which allow to expand any Feynman graph in coordinate
space in a set of {\em basic functions} and at the same time 
determine the (universal) renormalization of 
these functions. {\em The renormalized graph is then defined as
the corresponding linear combination of the
renormalized basic functions}. 
There are two types of rules, those stating how to
manipulate the singular diagrams and those fixing the
renormalization of the basic functions. The former
establish that all the algebra (including Dirac algebra in
four dimensions) must be simplified first, treating the singular 
expressions as if they were well-defined. This includes contracting 
all possible Lorentz indices and using the
Leibniz rule for derivatives. After this, the Feynman graph 
is written as a linear combination of total (external) derivatives of 
basic functions. These are products of propagators with all (internal)
derivatives acting only on the last propagator:
\bea
  \lefteqn{\F_{m_1m_2\dots m_n}^{(n)}[\OO^{(r)}](z_1,z_2,\dots,z_{n-1}) 
  \equiv} && \nn
  && \prop_{m_1}(z_1) \prop_{m_2}(z_2) \cdots \OO^{(r)z_1}
  \prop_{m_n}(z_1+z_2+\cdots +z_{n-1}) \, ,
\eea
where $\prop_{m_i}(z_i)=\frac{1}{4\pi^2} \frac{m_i K_1(m_i
z_i)}{z_i}$, with $K_1$ a modified Bessel function~\cite{Abramowitz},
is the propagator of a particle of mass $m_i$\footnote{
The expression of the propagator reduces to 
$\prop(z_i)=\frac{1}{4\pi^2} \frac{1}{z_i^2}$
in the massless case.} and
$\OO^{(r)z_1}$ is a differential operator of order $r$ acting on the
variable $z_1$. $z_i=x_i-x_{i+1}, ~i=1,\dots,n-1$ are
the coordinate differences and we work in four-dimensional 
euclidean space.
The rules of the second type 
reduce the degree of singularity (changing singular
expressions by derivatives of less singular ones, which are prescribed
to act formally by parts on test functions) and extend
mathematical identities among tempered distributions to arbitrary
functions. 
\begin{table}
\begin{flushleft}
  \begin{displaymath}
  \begin{array}{|c|c|c|c|}
    \hline 
      \makebox[3cm]{}& \makebox[3cm]{} & \makebox[3cm]{$\A_m[1]$}
       & \makebox[3cm]{$\A_m[\d_\mu]$} \\
    \hline 
     \B_{m_1m_2}[1] & \B_{m_1m_2}[\d_\mu] 
      & \B_{m_1m_2}[\Box] & \\
      & & \B_{m_1m_2}[\d_\mu\d_\nu] & \\
    \hline 
      \T_{m_1m_2m_3}[\Box]  & 
      \T_{m_1m_2m_3}[\Box\d_\mu] & & \\
      \T_{m_1m_2m_3}[\d_\mu\d_\nu] & 
      \T_{m_1m_2m_3}[\d_\mu\d_\nu\d_\rho] & & \\
    \hline 
      \Q_{m_1m_2m_3m_4}[\Box\Box] & & & \\
      \Q_{m_1m_2m_3m_4}[\Box \d_\mu\d_\nu] 
      & & &\\
      \Q_{m_1m_2m_3m_4}[\d_\mu\d_\nu\d_\rho\d_\sigma] & & & \\
    \hline
  \end{array}
  \end{displaymath}
\end{flushleft}
\caption{Singular basic functions for renormalizable gauge theories in
four dimensions in the Feynman gauge. Lines are ordered according to 
the number of propagators and columns according to the degree of 
singularity. The function $\A$, that appears in
tadpoles, is defined as $\A_m[\OO]=\OO \prop_m(z) \delta(z)$.}
\end{table}
In four-dimensional renormalizable theories 
(in the Feynman gauge) there are only thirteen
singular basic functions. 
They can have one (A), two (B), three (T) or four (Q) propagators,
corresponding to the lines in Table~1, and 
the singular behaviour at
coincident points can be of order $z^{-4}$ (logarithmic), $z^{-5}$
(linear), $z^{-6}$ (quadratic) and $z^{-7}$ (cubic), corresponding
to the columns in that table. A cubic singular behaviour
only occurs in tadpole diagrams that vanish for symmetry reasons
in usual theories. 
A detailed description of the CDR rules and the renormalized
expressions of
these singular basic functions in both coordinate and momentum space
are given in Ref.~\cite{techniques}.
With those expressions and the decomposition described above one can
systematically calculate any one-loop amplitude. A computer program
performing all operations automatically is available. Its description
can be found in Ref~\cite{program}.

\section{The gluon selfenergy}
To show how the method works, let us evaluate the gluon selfenergy 
in QCD. We shall first do the calculation in coordinate space, 
and afterwards perform a Fourier transform to obtain the corresponding
momentum space expression. 
To simplify the resulting expressions, we shall consider the case of 
massless quarks, although the massive case is analogous. 
The lagrangian and Feynman rules in coordinate space can be found in 
Ref.~\cite{QCD}.
\begin{figure}[h]
\begin{center}
\epsfxsize=10cm
\epsfbox{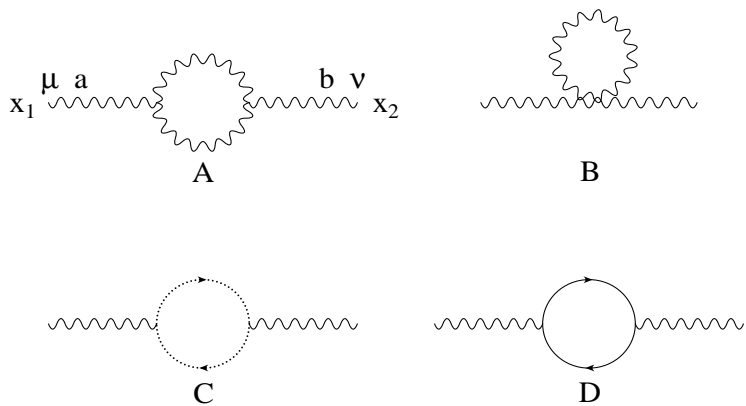}
\end{center}
\caption{Feynman diagrams contributing to
the gluon selfenergy.}
\end{figure}
The four contributing diagrams are depicted in Fig.~1.
Their ``bare'' expressions in coordinate space, 
in the Feynman gauge, are
\bea
\Pi_{\mu\nu}^{ab\,(A)} & = &
  \frac{1}{2} g^2 N_c \delta^{ab} \prop(x) \left[
  \delta_{\mu\sigma} (D_\rho-\d_\rho) + 
  \delta_{\sigma\rho} (\d_\mu-\stackrel{\leftarrow}{\d}_\mu) +
  \delta_{\rho\mu} (\stackrel{\leftarrow}{\d}_\sigma-D_\sigma) \right]
  \nn
  && \mbox{} 
  \left[ \delta_{\sigma\nu} (\d_\rho-D_\rho) +
  \delta_{\nu\rho} (D_\sigma-\stackrel{\leftarrow}{\d}_\sigma) +
  \delta_{\rho\sigma} (\stackrel{\leftarrow}{\d}_\nu-\d_\nu) \right] 
  \prop(x) \, , \label{pia} \\
\Pi_{\mu\nu}^{ab\,(B)} & = &
  -6 g^2 N_c \delta^{ab} \delta_{\mu\nu} \prop(x) \delta(x) \, , 
  \label{pib} \\
\Pi_{\mu\nu}^{ab\,(C)} & = &
  - g^2 N_c \delta^{ab} \d_\mu \prop(x) \d_\nu \prop(x) \, ,
  \label{pic} \\
\Pi_{\mu\nu}^{ab\,(D)} & = &
  - \frac{1}{2} g^2 N_f \delta^{ab}
  {\mathrm Tr} [\dsl \prop(x) \gamma_\mu \dsl \prop(x) \gamma_\nu] \, ,
  \label{pid} 
\eea
where $x\equiv z_1=x_1-x_2$ is the
coordinate difference, $N_c$ the number of colours and $N_f$ the number 
of quark flavors. The partial derivatives 
$\d_\alpha=\frac{\d}{\d x_\alpha}$ act on the
next propagator, while $\stackrel{\leftarrow}{\d}_\alpha$ act on the
previous one.
$D_\alpha$ are partial derivatives acting on the
external gluons. Using the Leibniz rule, they reduce to minus total
derivatives of the amputated expression.
We first expand Eqs.~(\ref{pia}--\ref{pid}) in basic functions:
\bea
\Pi_{\mu\nu}^{ab\,(A)} & = &
  \frac{1}{2} g^2 N_c \delta^{ab} \left\{
  (2\d_\mu\d_\nu-5\delta_{\mu\nu}\Box) \B[1]  
  + 5 (\d_\mu\B[\d_\nu]+\d_\nu\B[\d_\mu]) \right. \nn
  && \left. \mbox{} + 
  2\delta_{\mu\nu} \d_\sigma \B[\d_\sigma]
  -10 \B[\d_\mu\d_\nu] - 2 \delta_{\mu\nu} \B[\Box] \right\} \, , 
  \label{piBa} \\
\Pi_{\mu\nu}^{ab\,(B)} & = &
  -6 g^2 N_c \delta^{ab} \delta_{\mu\nu} \A[1] \, , 
  \label{piBb} \\
\Pi_{\mu\nu}^{ab\,(C)} & = &
  - g^2 N_c \delta^{ab} \left\{ \d_\mu \B[\d_\nu]-\B[\d_\mu\d_\nu])
  \right\}  \, , \label{piBc} \\
\Pi_{\mu\nu}^{ab\,(D)} & = &
  - 2 g^2 N_f \delta^{ab} \left\{
  \d_\mu \B[\d_\nu] + \d_\nu \B[\d_\mu] - 
  \delta_{\mu\nu} \d_\sigma \B[\d_\sigma] - 
  2\B[\d_\mu\d_\nu]  \right. \nn
  && \left. \mbox{} +
  \delta_{\mu\nu} \B[\Box] \right\} \, .
  \label{piBd}
\eea
Then, we replace the basic functions by their renormalized
expressions in Table~2 of Ref.~\cite{techniques}, namely
\bea
\A^R[1] & = & 0 \, , \\
\B^R[1] & = & - \frac{1}{64\pi^4} \Box 
  \frac{\log x^2 M^2}{x^2} \, , \\ 
\B^R[\d_\mu] &=&\frac{1}{2} \d_\mu \B^R[1] \, , \\
\B^R[\Box] & = & 0 \, ,\\
\B^R[\d_\mu\d_\nu] & = & \frac{1}{3} (\d_\mu\d_\nu
  -\frac{1}{4}\delta_{\mu\nu}\Box) \B^R[1]
  + \frac{1}{288\pi^2}(\d_\mu\d_\nu-\delta_{\mu\nu} \Box)
  \delta(x) \, ,
\eea
where the derivatives
are prescribed to act formally by parts and
$M$ is the renormalization scale.
The result for the gluon selfenergy is the sum of 
Eqs.~(\ref{piBa}--\ref{piBd})
with $\A[1]$ and $\B[\OO]$ replaced by $\A^R[1]$ and $\B^R[\OO]$:
\bea
  \Pi_{\mu\nu}^{ab\,R} & = &- \frac{1}{144\pi^2} g^2 \delta^{ab} \,
  (\d_\mu\d_\nu-\delta_{\mu\nu} \Box) \nn
  && \mbox{} 
  \left[ (15N_c-6N_f) \frac{1}{4\pi^2} \Box
  \frac{\log x^2M^2}{x^2} + (2N_c-2N_f) \delta(x) \right] \, .
\eea
The result is transverse ($\d_\mu \Pi_{\mu\nu}^{ab\,R} = 0$),
as required by gauge invariance~\cite{ST}.

$\Pi^{ab \, R}_{\mu\nu}$ is less singular than the bare
$\Pi^{ab}_{\mu\nu}$ and admits
a finite Fourier transform in four dimensions:
\be
  \hat{\Pi}^{ab \, R}_{\mu\nu}(p) = \int {\mathrm d}^4 x \,
  e^{ix \cdot p} \Pi^{ab \, R}_{\mu\nu}(x) \, .
\ee
This momentum space gluon selfenergy is easily obtained acting with
the external derivatives by parts on the exponential, as prescribed
by DR, and Fourier transforming $\frac{\log x^2M^2}{x^2}$~\cite{FJL}
and $\delta(x)$:
\bea
  \hat{\Pi}^{ab \, R}_{\mu\nu} &=& -\frac{1}{144\pi^2} g^2 \delta^{ab}
  (p_\mu p_\nu-\delta_{\mu\nu} p^2) \, \nn
  && \mbox{} \times
  \left[ (15N_c-6N_f) \log \frac{\hat{M}^2}{p^2} - 2N_c + 2N_f
  \right] \, ,
\eea
where $\hat{M}=2M/\gamma_E$ and $\gamma_E=1.781\dots$ is Euler's
constant. In more involved cases including masses one can use
the Fourier transforms of basic functions in Appendix~B of 
Ref.~\cite{techniques}.

Alternatively, one can do all the computations in momentum space, 
once the
Fourier transforms of basic functions are known.
Every step in the calculation in coordinate space can be translated
into momentum space.

\section{Anomalies}
When dealing with singular expressions, anomalies can appear at the quantum
level. In some cases (spurious anomalies) the symmetry
may be restored with finite local counterterms, while in others the
anomaly cannot be avoided and has physical relevance. This
is the case of the ABJ triangular anomaly~\cite{ABJ}.
In general, the source of anomalies
in CDR is the non-commutation of
renormalization with the contraction of Lorentz indices, 
which comes from the fact that the rules we have imposed 
are incompatible with this operation~\cite{techniques,CDR,polonia}.
Let us briefly discuss the case of the ABJ anomaly in QED. 
In the one-loop triangular diagram of one axial and two vector 
currents, the trace of six Dirac matrices and one $\gamma_5$ appears.
Although this trace is unique in four dimensions, it can be written
in different ways that lead to different expansions in (singular) 
basic functions (each of them with different contractions of the
internal derivatives). The final results for different expansions
differ by finite pieces, and in all cases 
at least one of the three
independent Ward identities is broken. We have checked that
the democratic form,
\bea
  \lefteqn{{\mathrm Tr}[\gamma_5 \gamma_\lambda\gamma_a\gamma_\nu
  \gamma_b\gamma_\mu\gamma_c] = 
  4 \left( \epsilon_{\lambda a\nu b} \delta_{\mu c}
  - \epsilon_{\lambda a \nu\mu} \delta_{bc} +
  \epsilon_{\lambda a \nu c} \delta_{b\mu} \right.} \nn
  && \mbox{} +
  \epsilon_{\lambda a b\mu} \delta_{\nu c}
  + \epsilon_{\lambda \nu bc} \delta_{a\mu} -
  \epsilon_{\lambda abc} \delta_{\mu\nu} 
  +
  \epsilon_{\lambda a\mu c} \delta_{\nu b}
  - \epsilon_{\lambda \nu\mu c} \delta_{ba} +
  \epsilon_{\lambda b\mu c} \delta_{a\nu} \nn
  && \left. \mbox{} +
  \epsilon_{a\nu b\mu} \delta_{\lambda c}
  + \epsilon_{a\nu\mu c} \delta_{\lambda b} -
  \epsilon_{a\nu bc} \delta_{\mu\lambda} 
  +
  \epsilon_{\nu b\mu c} \delta_{\lambda a}
  + \epsilon_{\nu b\mu \lambda} \delta_{ca} -
  \epsilon_{ab\mu c} \delta_{\lambda\nu} \right)  ,
\eea
preserves the vector Ward identities, giving the correct anomaly in
the axial current~\cite{CDR}. 
A detailed discussion of anomalies within constrained differential
renormalization will be presented elsewhere.

\section{Conclusions}
Constrained differential renormalization is a new version of DR which
renders renormalized amplitudes fulfilling automatically the Ward
identities of symmetric theories. It has been explicitly shown to work
at one loop for renormalizable gauge theories. 
Here we have verified that
the one-loop gluon selfenergy renormalized with this method satisfies 
the transversality condition implied by gauge invariance, and have
briefly discussed how the ABJ triangular anomaly is recovered.
One must also wonder about higher order calculations. They require
generalizing the renormalization rules (in particular, the treatment
of integrated internal points must be specified) as well as the 
set of basic functions. This issue is under study.

\section{Acknowledgements}
FA thanks the organizers for a pleasant meeting. This work has been 
supported by CICYT, under contract number AEN96-1672, and by Junta
de Andaluc\'{\i}a, FQM101. MPV thanks Ministerio de Educaci\'on y
Cultura for financial support.


\begin{thebibliography}{99}
\bibitem{FJL} D.Z. Freedman, K. Johnson and J.I. Latorre, 
    \np{B371}{92}{353}.
\bibitem{techniques} F. del Aguila, A. Culatti,
    R. Mun\~oz Tapia and M. P\'erez-Victoria, MIT-CTP-2705,
    UG-FT-86/98, hep-ph/9806451.
\bibitem{CDR} F. del Aguila, A. Culatti, R. Mu\~noz Tapia and
    M. P\'erez-Victoria, \pl{B419}{98}{263}.
\bibitem{polonia} F. del Aguila and M. P\'erez-Victoria, 
    \vj{Acta Phys. Polon.}{B28}{97}{2279}.
\bibitem{QCD} M. P\'erez-Victoria, UG-FT-89/98, hep-th/9808071.
\bibitem{program} T. Hahn and M. P\'erez-Victoria, 
    UG-FT-87/98, KA-TP-7-1998, hep-ph/9807565.
\bibitem{ST} A.A. Slavnov, \vj{Theor. and Math. Phys.}{10}{72}{99};
     J.C. Taylor, \np{B33}{71}{436}.
\bibitem{Abramowitz} M. Abramowitz and I.A. Stegun,
    {\em Handbook of Mathematical Functions} (Dover, New York, 1972).
\bibitem{ABJ} S. Adler, \pr{177}{69}{2426}; 
     J.S. Bell and R. Jackiw, \vj{Nuovo Cimento}{51}{69}{47}.
\end{thebibliography}
\end{document}